\documentclass[aps,prc,twocolumn,showpacs,groupedaddress,floatfix]{revtex4}

\usepackage{graphicx}
\usepackage{dcolumn}
\usepackage{amsfonts}

\begin{document}

\title{Dynamical deformation effects in subbarrier fusion of $^{64}$Ni+$^{132}$Sn}


\author{A.S. Umar and V.E. Oberacker}
\affiliation{Department of Physics and Astronomy, Vanderbilt University, Nashville, Tennessee 37235, USA}

\date{\today}


\begin{abstract}
We show that dynamical deformation effects play an important role in fusion reactions involving
the $^{64}$Ni nucleus, in particular the $^{64}$Ni+$^{132}$Sn system. We calculate fully microscopic
interaction potentials and the corresponding subbarrier fusion cross sections.
\end{abstract}
\maketitle


Recently observed enhanced subbarrier fusion cross sections for the neutron rich $^{64}$Ni+$^{132}$Sn
system~\cite{Li03} has further invigorated the research in low-energy nuclear reactions involving
exotic nuclei.
For this system fusion cross-sections were measured in the energy range 142~MeV~$\le E_{\mathrm{c.m.}} \le$~176~MeV.
In particular, it was found that fission is negligible for $E_{\mathrm{c.m.}}\le$~160~MeV
and therefore the evaporation residue cross-sections have been taken as fusion cross sections.
The enhancement of subbarrier fusion was deduced from comparison with a barrier penetration
calculation, using a phenomenological Woods-Saxon interaction potential whose parameters were
fitted to reproduce the evaporation residue cross sections
for the $^{64}$Ni+$^{124}$Sn system. Similarly, sophisticated coupled-channel calculations~\cite{BT98,Esb04},
which are known to enhance the fusion cross sections by considering coupling to various excitation
channels and neutron transfer, have significantly underestimated the subbarrier fusion cross sections for the $^{64}$Ni+$^{132}$Sn system~\cite{Li03}.

In general, the fusion cross sections depend on the interaction potential and form factors in the
vicinity of the Coulomb barrier. These are expected to be modified during the collision due
to dynamical effects. In addition, experiments on subbarrier
fusion have demonstrated a strong dependence of the total fusion cross section
on nuclear deformation~\cite{SE78}. The dependence on nuclear orientation has received
particular attention for the formation of heavy and superheavy elements~\cite{KH02}
and various entrance channel models have been developed to predict its role in
enhancing or diminishing the probability for fusion~\cite{SC04,UO06c}.

Recently, we have developed a new approach for calculating heavy-ion interaction
potentials which incorporates all of the dynamical entrance channel effects included in the
time-dependent Hartree-Fock (TDHF) description of the collision process~\cite{UO06b}.
These effects include the neck formation, particle exchange, internal excitations,
and deformation effects to all order, as well as the effect of nuclear alignment
for deformed systems~\cite{UO06b,UO06c}. The method is based on the TDHF
evolution of the nuclear system coupled with density-constrained Hartree-Fock
(DCHF) calculations~\cite{CR85,US85} to obtain the interaction potential,
given by
\begin{equation}
V(R)=E_{\mathrm{DC}}(R)-E_{A_{1}}-E_{A_{2}}\;.
\label{eq:vr}
\end{equation}
The potential deduced from Eq.~(\ref{eq:vr}) contains
{\it no parameters} and {\it no normalization}.
Given an effective interaction, such as the Skyrme force, $V(R)$
can be constructed by performing a TDHF evolution and minimizing
the energy at certain times to obtain
$E_{\mathrm{DC}}(R)$, while the nuclear binding energies $E_{A_{1}}$ and $E_{A_{2}}$ are the
results of a static Hartree-Fock calculation with the same effective
interaction~\cite{UO06b}.

We have carried out a number of TDHF calculations with accompanying
density constraint calculations to compute $V(R)$ given by Eq.~(\ref{eq:vr}).
A detailed description of our new three-dimensional unrestricted TDHF code
has recently been published in Ref~\cite{UO06a}.
For the effective interaction we have used the Skyrme SLy5 force~\cite{CB98}
including all of the time-odd terms.
In our case the $^{64}$Ni nucleus is essentially oblate with a small mix of
triaxiality, having a quadrupole moment of -0.45~b. This is
also confirmed by other calculations~\cite{DS04,LR99} and suggested by
experiments~\cite{VA90}.
\begin{figure}[!htb]
\begin{center}
\includegraphics*[scale=0.40]{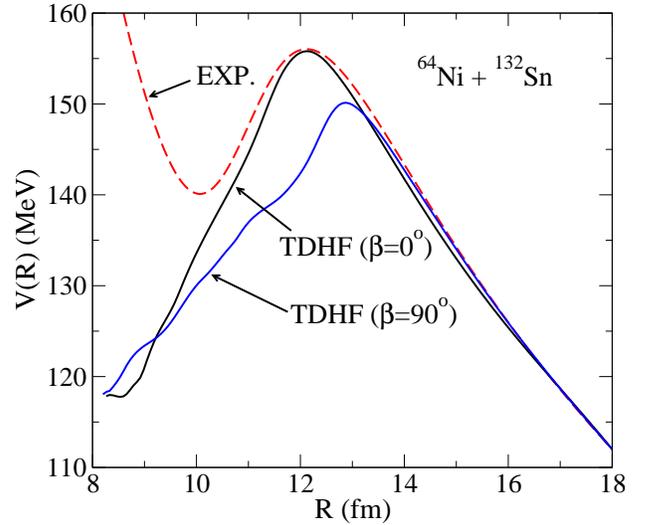}
\caption{\label{fig:vnr} (Color online) Internuclear potential obtained from Eq.~(\protect\ref{eq:vr})
for the head-on collision of the $^{64}$Ni+$^{132}$Sn system
at $E_{\mathrm{c.m.}}=158$~MeV. The dashed line shows the empirical Woods-Saxon potential used in Ref.~\protect\cite{Li03}.}
\end{center}
\end{figure}
TDHF calculations were initialized at $E_{\mathrm{c.m.}}=158$~MeV. In Ref.~\cite{UO06b} we
have shown that the calculation of the potential barrier is not sensitive to the choice
of the TDHF initialization energy, the only difference being a slightly lower potential well
for lower energies.
In Fig.~\ref{fig:vnr} we
show the results obtained for the interaction potential as well as the empirical potential barrier
used in Ref.~\cite{Li03}. The angle $\beta$ indicates the orientation of the symmetry
axis. In the case of $\beta=0^{\circ}$ the symmetry axis of the oblate $^{64}$Ni is aligned with the
collision axis and for $\beta=90^{\circ}$ the symmetry axis is perpendicular to the collision
axis. For the case of parallel orientation the calculated barrier is almost exactly
the same as the one used in Ref.~\cite{Li03}, having a barrier height of 155.8~MeV. The
difference for smaller $R$ values is due to the use of the point Coulomb interaction
in the model calculation, which is unphysical when nuclei overlap. The same argument
applies to small differences at large $R$ values, since the Coulomb interaction is
slightly different due to the deformed Ni nucleus.
We would like to
emphasize again that our calculations do not contain any parameters or normalization.
On the other hand, the barrier corresponding to the perpendicular alignment is considerably
lower, peaking at 150.1~MeV, and has a narrower width.

The physical picture which emerges from these calculations is that for center-of-mass
energies in the range 150.1-155.8~MeV the fusion cross section would be dominated by the channel above
the lower barrier since the contribution via tunneling through the higher barrier will be
substantially smaller. Similarly, for energies below 150.1~MeV transmission through the lower
barrier will produce the dominant contribution. Of course, for energies above 155.8~MeV
both barriers will contribute. As a result, the only data point which is truly
subbarrier is the lowest energy point.
\begin{figure}[!htb]
\begin{center}
\includegraphics*[scale=0.40]{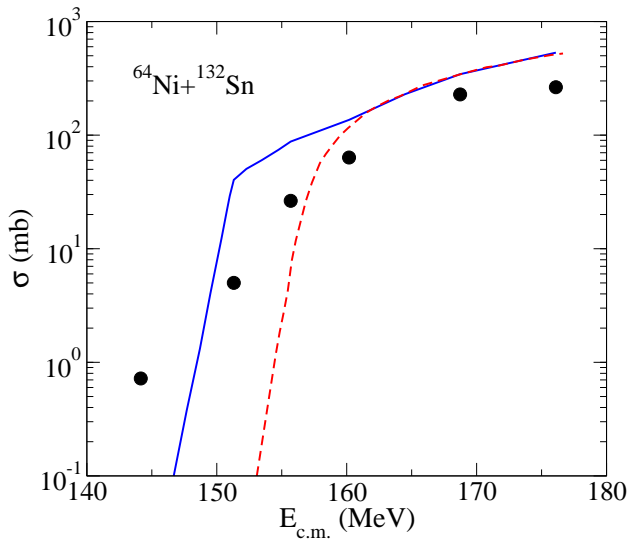}
\caption{\label{fig:sigma} (Color online) Fusion cross sections obtained for the $^{64}$Ni+$^{132}$Sn system
using the microscopically calculated potentials discussed in the manuscript.
Also shown (circles) are the experimental values from Ref.~\protect\cite{Li03}.}
\end{center}
\end{figure}
In Fig.~\ref{fig:sigma} we show the calculated fusion cross sections as a function of the
center-of-mass energy. We have used a simple WKB approach to calculate the cross section
for the lowest energy point and the parabolic approximation via the Wong formula~\cite{Wong}
for the higher energy points.
Also shown are the corresponding experimental values (circles) as well as the barrier
peneration model results (dashed line) from Ref.~\cite{Li03}. As
anticipated, the calculated cross sections, with the exception of the lowest energy data
point, are now above the corresponding experimental values. There are a number of reasons for
over-predicting the data at higher energies.
The first reason is the alignment probability
of the deformed nucleus, which could be calculated with the availability of the excitation
spectrum for the $^{64}$Ni nucleus~\cite{UO06c}. Physically, one expects a distribution of
barriers starting from the lowest barrier and approaching the highest barrier. The second factor
is the quality of the parabolic approximation used in the Wong formula. It is well known that
the rising Coulomb tail of barriers cannot be properly accounted for by a single parabola,
thus resulting in a somewhat thinner barrier and a larger fusion cross section. In addition,
for energies above 160~MeV the fission channel opens up.
Finally, despite improving the lowest energy cross section by many orders of magnitude we find
a cross section of 0.0035~mb, which is still a factor of 200 lower than the experimental
value.

In conclusion, we have performed microscopic calculations of the interaction
potentials for the $^{64}$Ni+$^{132}$Sn system. We observe that dynamical
deformation effects play a very significant role in the calculation of fusion
cross sections. This observation further underscores the necessity of
detailed structure information for neutron and proton rich systems for the
better description of fusion cross sections involving these nuclei. Availability
of detailed structure data for the $^{64}$Ni nucleus may help explain the
discrepancy for the lowest energy point.

This work has been supported by the U.S. Department of Energy under grant No.
DE-FG02-96ER40963 with Vanderbilt University. Some of the numerical calculations
were carried out at the IBM-RS/6000 SP supercomputer of the National Energy Research
Scientific Computing Center which is supported by the Office of Science of the
U.S. Department of Energy.


\begin{thebibliography}{99}

\bibitem{Li03} J. F. Liang {\it et al.}, Phys. Rev. Lett. {\bf 91}, 152701 (2003);
                                         Phys. Rev. Lett. {\bf 96}, 029903(E) (2006).

\bibitem{BT98} A. B. Balantekin and N. Takigawa, Rev. Mod. Phys. {\bf 70}, 77 (1998).

\bibitem{Esb04} H. Esbensen, Prog. Theor. Phys. Suppl. {\bf 154}, 11 (2004).

\bibitem{SE78} R. G. Stokstad, Y. Eisen, S. Kaplanis, D. Pelte, U. Smilansky, and
               I. Tserruya, Phys. Rev. Lett. {\bf 41}, 465 (1978).

\bibitem{KH02} K. Nishio, {\it et al.}, J. Nucl. Radiochemical Sci., {\bf 3}, 89 (2002).

\bibitem{SC04} C. Simenel, Ph. Chomaz, and G. de France, Phys. Rev. Lett. {\bf 93}, 102701-1 (2004).

\bibitem{UO06c} A. S. Umar and V. E. Oberacker, Phys. Rev. C {\bf 74}, 024606 (2006).

\bibitem{UO06b} A. S. Umar and V. E. Oberacker, Phys. Rev. C {\bf 74}, 021601(R) (2006).

\bibitem{CR85} R. Y. Cusson, P. -G. Reinhard, M. R. Strayer, J. A. Maruhn, and W. Greiner,
               Z. Phys. A {\bf 320}, 475 (1985).

\bibitem{US85} A. S. Umar, M. R. Strayer, R. Y. Cusson, P. -G. Reinhard,
               and D. A. Bromley, Phys. Rev. C {\bf 32}, 172 (1985).

\bibitem{UO06a} A. S. Umar and V. E. Oberacker, Phys. Rev. C {\bf 73}, 054607 (2006).

\bibitem{CB98} E. Chabanat, P. Bonche, P. Haensel, J. Meyer and R. Schaeffer,
               Nucl. Phys. {\bf A635}, 231 (1998); Nucl. Phys. {\bf A643}, 441 (1998).

\bibitem{DS04} J. Dobaczewski, M. V. Stoitsov, and W. Nazarewicz, {\it Skyrme-HFB deformed nuclear mass table},
               ed. R. Bijker, R.F. Casten, and A. Frank, (AIP, New York, 2004), {\bf 726}, 51 (2004).

\bibitem{LR99} G. A. Lalazissis, S. Raman, and P. Ring, At. Data Nucl. Data Tables {\bf 71}, 1 (1999).

\bibitem{VA90} J. J. Vega, E. F. Aguilera, G. Murillo, J. J. Kolata, A. Morsad, and X. J. Kong,
               Phys. Rev. C {\bf 42}, 947 (1990).

\bibitem{Wong} C. Y. Wong, Phys. Rev. Lett. {\bf 31}, 766 (1973).
\end{thebibliography}
\end{document}